\documentclass{aastex631}
\usepackage{bm,amsmath,amssymb,amsfonts,mathrsfs,enumitem}
\newcommand{\dd}{{\rm d}}
\newcommand{\bx}{\boldsymbol{x}}

\newcommand{\ba}{\boldsymbol{a}}
\newcommand{\bvv}{\boldsymbol{v}}

\newcommand{\bq}{\boldsymbol{q}}

\newcommand{\bn}{\boldsymbol{n}}

\newcommand{\bp}{\boldsymbol{p}}

\newcommand{\muas}{ \mu{\rm as}}

\begin{document}

\title{Dark Photon Dark Matter and Low-Frequency Gravitational Wave Detection with Gaia-like Astrometry}
	\author{Haipeng An}
	\email{anhp@mail.tsinghua.edu.cn}
	\affiliation{Department of Physics, Tsinghua University, Beijing 100084, China}
	\affiliation{Center for High Energy Physics, Tsinghua University, Beijing 100084, China}
	\affiliation{Center for High Energy Physics, Peking University, Beijing 100871, China}	
 \author{Tingyu Li}
 \email{lity22@mails.tsinghua.edu.cn}
	\affiliation{Department of Physics, Tsinghua University, Beijing 100084, China}
\author{Jing Shu}
\email{jshu@pku.edu.cn}
\affiliation{School of Physics and State Key Laboratory of Nuclear Physics and Technology, 
	Peking University, Beijing 100871, China}
\affiliation{Center for High Energy Physics, Peking University, Beijing 100871, China}
\affiliation{Beijing Laser Acceleration Innovation Center, Huairou, Beijing, 101400, China}
\author{Xin Wang}
\email{xin.wang@soton.ac.uk}
\affiliation{School of Physics and Astronomy, University of Southampton, Southampton SO17 1BJ, United Kingdom} 
\affiliation{School of Physics and State Key Laboratory of Nuclear Physics and Technology, 
	Peking University, Beijing 100871, China}
\author{Xiao Xue}
\email{xiao.xue@desy.de}
\affiliation{II. Institute of Theoretical Physics, Universit\"{a}t Hamburg, 22761 Hamburg, Germany}
\affiliation{Deutsches Elektronen-Synchrotron DESY, Notkestr. 85, 22607 Hamburg, Germany}
\author{Yue Zhao}
\email{zhaoyue@physics.utah.edu}
\affiliation{Department of Physics and Astronomy, University of Utah, Salt Lake City, UT 84112, USA}
\begin{abstract}
Astrometric surveys offer us a method to search for elusive cosmic signatures, such as ultralight dark photon dark matter and gravitational waves, by observing the temporal change of stars' apparent location. The detection capabilities of such surveys rapidly decrease at low frequencies, because the signals become hardly distinguishable from the background motion of stars. In this work, we find that the background motion can be well described by a linear model over time, based on which we propose a linear background subtraction scheme. Compared to the conventional quadratic subtraction, the advantage of linear subtraction emerges within the frequency range below $6 \times 10^{-9}~{\rm Hz}$. Taking dark photons with purely gravitational interactions, dark photons with additional $U(1)_{B}$ or $U(1)_{B-L}$ gauge interactions, and low-frequency gravitational waves as examples, we illustrate that the linear subtraction scheme can result in an enhancement of more than one order of magnitude in the exclusion limits of Gaia-like experiments in the low-frequency range.\\
\\
preprint: DESY-24-106
\end{abstract}
\section{Introduction}
The development of astrophysical observation methods has revolutionized our understanding of the cosmos, enabling us to probe subtle effects induced by new astrophysical sources or new physics with unprecedented precision. Recently, several Pulsar Timing Array (PTA) collaborations reported first evidence for the stochastic gravitational wave (GW) background at nano-Hertz (nHz) frequencies
~(\cite{Allen:1996vm,NANOGrav:2023gor, EPTA:2023fyk, Reardon:2023gzh, Xu:2023wog}), which may arise from a population of inspiraling supermassive black hole binaries~(\cite{Sesana:2008mz}).

Apart from the GW signals, PTA experiments can also probe several new physics models. One example would be the ultralight bosonic dark matter (\cite{Schive:2014dra,Hui:2016ltb})
which has the potential to solve the so called small-scale problems (\cite{Weinberg:2013aya}) of the cold dark matter model. Ultralight dark photon dark matter (DPDM) assumes that dark matter is composed of vector fields~(\cite{Adshead:2021kvl,Amin:2022pzv}) with a mass around $10^{-22}~{\rm eV}$.
Due to its wave nature, the field value of the ultralight DPDM oscillates in an approximately coherent manner. Therefore, the ultralight DPDM can induce nontrivial oscillations of the metric due to the pure gravitational effects~(\cite{Khmelnitsky:2013lxt,Nomura:2019cvc,Chen:2022kzv,Yu:2024enm}), or impose external oscillating forces on the test objects via additional U(1) gauge interactions~(\cite{Pierce:2018xmy}), both of which can modify the pulse arrival time~(\cite{Porayko:2018sfa,PPTA:2021uzb,NANOGrav:2023hvm, Nelson:2011sf, PPTA:2022eul, Xia:2023hov,Chowdhury:2023xvy}).  

The GW or DPDM background can also result in the deflection of light emitted from the astrophysical objects to the Earth due to the following two distinct effects:
\begin{itemize}[leftmargin=*]
    \item GWs or DPDM with purely gravitational interactions can induce metric perturbations, which can affect the trajectory of light emitted from the sources~(\cite{braginsky1990propagation,Pyne:1995iy,kaiser1997bending,Kopeikin:1998ts,Book:2010pf}). As a result, the apparent positions of stars we observe slightly deviate from their initial positions. Such displacements depend on metric perturbations at both the Earth and stars, while the Earth term should be dominant, as we will explain later. It should also be mentioned that,
    unlike the deflection induced by ultralight scalar dark matter, which is suppressed due to an approximate spherical symmetry, the deflection caused by DPDM is generally not suppressed~(\cite{Chen:2022kzv}).
    \item DPDM coupling to the baryon number ${\rm U}(1)_B$ or the baryon number minus lepton number ${\rm U}(1)_{B-L}$ can lead to the aberration effect, which also changes the apparent position of stars~(\cite{Guo:2019qgs}).
\end{itemize}
By precisely measuring the variation of stars' apparent positions, it is hopeful of capturing the traces of GW and DPDM signals in the Gaia mission~(\cite{Gaia:2016zol}) and upcoming astrometric surveys like Roman~(\cite{Wang:2020pmf,Wang:2022sxn,Haiman:2023drc,Pardo:2023cag}) and Theia~(\cite{Theia:2017xtk,Malbet:2021rgr,Garcia-Bellido:2021zgu}). Gaia observes positions, proper motions, and parallaxes of more than $10^9$ objects with unprecedented accuracy~(\cite{Gaia:2016zol}). The sensitive frequency of Gaia is determined by its survey lifetime $T^{}_{\rm obs} = 10~{\rm year}$ and observational cadence $\Delta t = 24~{\rm day}$, typically, $10^{-9}~{\rm Hz} \lesssim f \lesssim 10^{-7}~{\rm Hz}$.  Hence Gaia can provide complementary detection capabilities to PTA experiments for nHz astrophysical sources~(\cite{ Moore:2017ity, Klioner:2017asb, Qin:2018yhy, OBeirne:2018slh, Mihaylov:2018uqm, Bini:2018fnv, Darling:2018hmc, Mihaylov:2019lft, Guo:2019qgs, Qin:2020hfy, Aoyama:2021xhj, Jaraba:2023djs, Liang:2023pbj,Caliskan:2023cqm}).

To extract the deflections to the apparent positions of stars caused by external signals, we need to subtract their intrinsic background motion stemming from the physical motion of stars with respect to the solar barycenter and the secular aberration due to the moving reference system. This intrinsic motion can induce large angular deflection far exceeding the resolution of Gaia. For the signals with frequencies $f \gtrsim 2/T_{\rm obs} \sim 6 \times 10^{-9}~{\rm Hz}$, the background subtraction essentially does not influence the sensitivity, since the oscillation pattern of GWs or DPDM is distinct from the background. 
Conversely, if $f \lesssim 6 \times 10^{-9}_{}~{\rm Hz}$, a half oscillating cycle would not be observable in the time-series data within the operational lifetime of Gaia. In such cases, the signal is also partially subtracted, thereby reducing the sensitivity of detecting low-frequency GW and DPDM signals. Analogous to PTAs, the quadratic model is commonly used to subtract the background proper motion in the existing literature~(\cite{Moore:2017ity, Guo:2019qgs, Wang:2020pmf, Wang:2022sxn}). However, a more comprehensive study of the background motion of stars and subtraction schemes should be implemented.

In this work, we propose a linear background subtraction scheme for the time-series data of apparent stellar positions in Gaia-like astrometric surveys. Specifically, we apply the least squares method to fit a linear function to the data, modeling the background noise. This fitted function is then subtracted from the data, effectively isolating the signal of interest. Upon analyzing the background motion of stars, we find that the impact on the apparent stellar position due to the constant acceleration from the galactic gravitational potential is negligibly small relative to Gaia's sensitivity. Meanwhile, the galactocentric acceleration of the solar barycenter mimics a linear term over time in the proper motion due to aberration. These observations serve as the primary motivation for our linear subtraction approach. Nevertheless, unresolvable binary systems could bring additional noise, which cannot be effectively subtracted. We provide concrete examples by calculating the astrometric deflection induced by DPDM with purely gravitational interactions while also revisiting scenarios involving DPDM with ${\rm U}(1)_B$ or ${\rm U}(1)_{B-L}$ gauge symmetry and GWs from binary systems. Subsequent numerical simulation using mock data reveals that our linear subtraction scheme significantly enhances Gaia's detection capabilities in the low-frequency range, improving sensitivity by more than an order of magnitude compared to the quadratic subtraction.

The layout of this paper is as follows. In Sec.~\ref{sec:background}, we illustrate the background motion of stars can be well described by a linear model over time. In Sec.~\ref{sec:deflection}, we calculate astrometric deflections induced by GWs and DPDM. The numerical simulation and results are presented in Sec.~\ref{sec:numerical}. We summarize our main conclusions in Sec.~\ref{sec:conclusion}.

\section{Background motion of the apparent stellar locations} \label{sec:background}
Astrometric objects exhibit background motion around the galactic center. Their motion is randomly distributed with a potentially non-zero expectation value depending on the star's sky location, due to the common motion of the galactic disk. We assume that the velocity $\bvv = (v_r, v_\theta,v_\phi)$ in spherical polar coordinates relative to the galactic center follows a Gaussian distribution
 \begin{align}
P(\bvv|\bn) = \prod_{i=1}^3\frac{1}{\sqrt{2\pi}\sigma_{v,i}} \exp\left(\frac{(v_i - v_{0,i}(\bn))^2 }{2\sigma_{v,i}^2}\right) \; ,
 \end{align}
where $\bn$ denotes the sky location of stars, $v_{0,i}(\bn)$ represents the expectation value of $\bvv$ in each direction, and $\sigma_{v,i}$ refers to the velocity dispersion. The anisotropy of the velocity dispersion has been observed by SDSS and Gaia. Approximately we have the relation $\sigma_{v,r}\simeq 2\sigma_{v,\theta/\phi}$ for stars with high metallicity, which may originate from a major merger event of a satellite galaxy between $8$ and $11$ Gyr ago~(\cite{belokurov2018co}). 
For qualitative discussions, we assume the uncertainty $\sigma_{v}\simeq 100\,{\rm km}/{\rm s} \simeq 1.01 \times 10^{-4}\,{\rm pc/year}$ for all stars on both directions on the sky sphere. At a typical distance of 1 kpc,\footnote{The peak distribution of distances between the stars and us observed by Gaia is $1 \sim 2$ kpc~(\cite{Vallenari:2866069}).}  the angular deflection observed over $10$ years reaches $\simeq 1.01 \times 10^{-6}\,{\rm rad}$, which significantly exceeds the single-exposure resolution $\sigma_{\rm ss} \simeq 100\,\mu{\rm as} = 4.85 \times 10^{-10}\,{\rm rad}$ of Gaia for a single star. It then becomes necessary to subtract the background proper motion from the data before searching for the GW or DPDM. 

We now turn our attention to the accelerations affecting stellar and the Earth's motion. Let us consider stars in two categories: isolated stars and the ones in binary or higher-order systems. For isolated stars, their velocities, in general, change over time. Hence, we should also estimate their accelerations, which mainly include contributions from two aspects:
\begin{itemize}[leftmargin=*]
    \item {\it The constant accelerations of stars}. The density gradient in the Milky Way could induce constant accelerations for stars. Unlike their proper motion, the constant accelerations of stars are predictable due to the galactic mass profile, indicating the uncertainty $\sigma_a$ should be much smaller than the center value of the star's acceleration, defined as $\ba_0$. Typical stars in the Milky Way experience an acceleration of $|\ba_0|\sim 10^{-10}\,{\rm m}/{\rm s}^2$~(\cite{silverwood2019stellar}),  which leads to a displacement of $|\ba_0|\Delta t^2/2 \simeq 1.6\times 10^{-10}$ pc over the 10-year observation. For a star located 1 kpc away from us, the corresponding angular deflection is about $0.03\,\muas$, significantly below the detection threshold. 
    
    \item {\it The constant acceleration of solar system barycenter}. Like other stellar systems, the velocity of our solar system is not precisely constant. Instead, it follows a curved orbit in the Galaxy. For the observer, the nonzero acceleration of the solar system barycenter gives rise to the aberration effect, which can be considered as extra (spurious) proper motion superimposed on stars~(\cite{Kopeikin:2005fc,klioner2021gaia})
    \begin{align}
    \frac{\dd\delta \bn}{\dd t} =  \ba_{\rm SSB} - (\ba_{\rm SSB}\cdot\bn)\bn \; ,
    \label{eq:aberra}
\end{align}
where $\delta\bn$ denotes the displacement in the apparent stellar position due to aberration, and $\ba_{\rm SSB}$ is the acceleration of the solar system barycenter (SSB). The estimated value of ${\bm a}_{\rm SSB}$ is $2.32\times 10^{-10}~{\rm m}/{\rm s}^{2}$, corresponding to a proper motion amplitude of $5.05~\mu{\rm as}/{\rm year}$~(\cite{klioner2021gaia}). From Eq.~(\ref{eq:aberra}), we realize that the displacement $\delta\bn$ is linear in time, as the aberration effect should be proportional to the velocity of the observer. Therefore, although the proper motion caused by ${\bm a}_{\rm SSB}$ is larger than Gaia's resolution, it can be subtracted by a function that is linearly dependent on time.

\end{itemize}

In short, the background motion of the apparent stellar locations is governed by the proper motion of stars, which can be significantly greater than the resolution of Gaia. The acceleration of stars is very minor and thus plays a negligible role. Moreover, the acceleration of the solar system barycenter yields spurious proper motion, which can be subtracted by a linear function in time. The above backgrounds can be eliminated through a linear subtraction scheme. 

Next, we investigate the significant background motion from unresolved binary or higher-order systems in the Galaxy.
In the practical numerical calculation, we employ a data compression technique akin to the one described in~\cite{Moore:2017ity, Klioner:2017asb}, since the vast number of objects ($\sim10^9$) in the full dataset of Gaia severely slows down the process of numerical simulation. Specifically, we compress the data of $10^5$ stars in the same sky area into one virtual star. For each star, the probability of being part of a binary or high-order system and having specific orbital parameters is statistically independent of the status of the other stars. After data compression, the astrometric data of $10^5$ stars in each sky area are added together. The background motion of individual stars in binary or higher-order systems manifests as a time-dependent  noise of the compressed virtual star. 

A survey of nearby ($<250$ pc) stellar objects finds $\sim 55\%$ solar-type stars are isolated, while the remaining $\sim 45\%$ stars are in binary or higher-order systems~(\cite{raghavan2010survey}). Here we consider the binary system consisting of two objects with masses $m_1$ and $m_2$ for simplicity. Following the Kepler's law, the orbital period $P_b$ has the following relation with the semi-major axis $a_b$
\begin{equation}
    a_b = \left[\frac{Gm_1(1+q)P_{ b}^2}{(2\pi)^2} \right]^{1/3} \; ,
\end{equation}
where $G$ is the gravitational constant and $q = m_2/m_1<1$. If both stars are observed by Gaia, the net contribution from the binary to the displacement of stars will approximate to
\begin{equation}
    r_{\rm net} \simeq a_b(1+e_b)\frac{1-q}{1+q} \; ,
\end{equation}
where $e_b$ is the orbit eccentricity. Then the net contribution will be fully canceled if $q=1$. 

\begin{figure}[t!]
    \centering
    \includegraphics[width=0.6\textwidth]{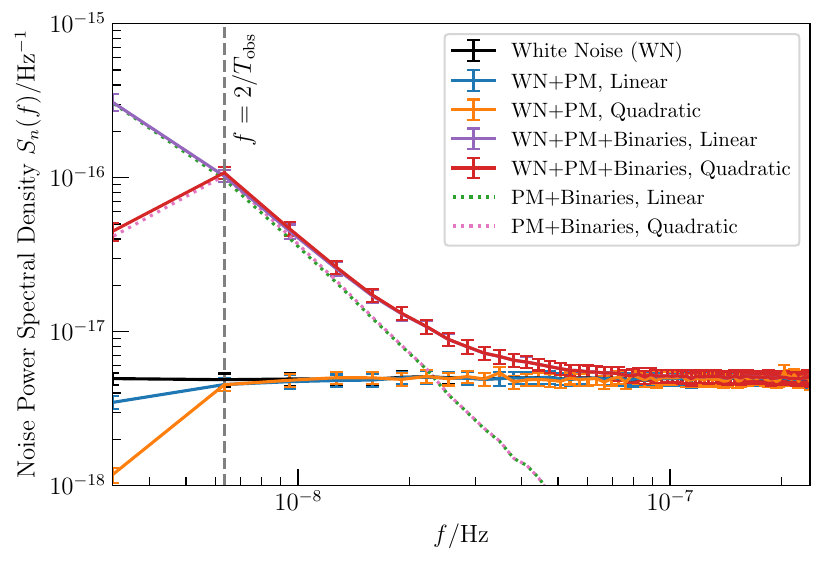}
    \caption{The average power spectral densities of the injected white noise (WN), proper motion (PM), and stellar binary motion, along with the associated error bars for each bin, obtained from $10^4$ simulations.  For clarity, the error bars have been magnified by {\bf a factor of 3} to improve visibility. Green and pink dashed lines correspond to the cases with WN removed. The vertical dashed line refers to $f = 2/T_{\rm obs}$ with $T_{\rm obs} = 10~{\rm year}$ being Gaia's observation period. The input Gaia operational parameters align with the discussion in Sec.~\ref{sec:numerical}.} 
    \label{fig:psd}
\end{figure}

For the $q < 1$ case, we illustrate by considering a $e_b=0$ binary composed of two stars with the total mass $M=m_1+m_2$. The net amplitude of the observed angular deflection in one direction is 
\begin{equation}
\begin{split}
A_{b} =  \frac{r_{\rm net} \lambda(\iota_b, \psi_b)}{\sqrt{2}D_b} 
 &=  6.10\times 10^{-9} {\rm rad} \left(\frac{1-q}{1+q}\right) \frac{\lambda(\iota_b, \psi_b)}{\sqrt{2}}\\
 & \times \left[\frac{P_b}{\rm year}\right]^{2/3} \left[\frac{D_b}{1{\rm kpc}}\right]^{-1}\left[\frac{M}{2M_{\odot}}\right]^{1/3} \; ,
\end{split}
\label{eq:Amp_binary}
\end{equation}
where $D_b$ is the distance between the observer and the binary barycenter and $M_{\odot}$ denotes the solar mass. Also,  $\lambda(\iota_b, \psi_b) = \cos\iota_b\sin\psi_b + \cos\psi_b$ with $\iota_b$ and $\psi_b$ being the inclination angle and the polarization angle of the stellar binary plane, respectively. Eq.~(\ref{eq:Amp_binary}) indicates binaries with $P_b \gtrsim 1\,{\rm year}$ and $D_b \sim 1\,{\rm kpc}$ may be resolved by Gaia. 
According to \cite{raghavan2010survey,simonetti2020statistical}, the period distribution approximately follows the log-Normal distribution
\begin{equation}
    P(\log_{10} P_b/{\rm day}) \simeq \frac{1}{2.28\sqrt{2\pi}}e^{ - \frac{\left(\log_{10} P_b/{\rm day}- 5.03\right)^2}{ 2\times 2.28^2}} \; ,\label{eq:Pb_distribution}
\end{equation}
which centers around a period of $293.6$ years. About $20\%$ stellar binaries have the orbital periods from $24$ days to $10$ years, which would induce an irreducible background for the $10$-year observation of Gaia mission.

We calculate the power spectral density (PSD) originating from binary stars $S_{\rm bin}(f)$ by implementing a Monte-Carlo simulation, where we assume $45\%$ of stars are in binaries. We use Eq. (\ref{eq:Amp_binary}) with $D_b=1$ kpc, while the orbital period is drawn from Eq.~(\ref{eq:Pb_distribution}). Furthermore, $\cos\iota_b$, $\psi_b$ and $q$ are treated as free parameters, randomly selected from uniform distributions within the intervals of $[-1,1)$, $[0,2\pi)$ and $[0.1,1)$, respectively. In each simulation, we generate astrometric data of $10^5$ stars, which are then compressed into one virtual star. We calculate the PSD of astrometric data of the virtual star. The final $S_{\rm bin}(f)$ is obtained by averaging the PSD over all simulations. 
The total noise PSD $S_n(f)$ is the combination of $S_{\rm bin}(f)$ and that from the instrumental noise $S_w(f)$, namely,
\begin{align}
    S_n(f) = S_w(f) + S_{\rm bin}(f) \; .\label{eq:psd_n}
\end{align}

We present the average noise power spectral density (PSD) in Fig.~\ref{fig:psd}, together with error bars of each bin. The uncertainties of PSDs estimated from the simulation are around $10^{-19}/{\rm Hz}^{-1}$. In the absence of the noise from stellar binaries, the linear/quadratic subtraction approach effectively reduces the noise from stellar proper motion, reproducing the PSD of pure white noise above $2/T_{\rm obs} \sim 6\times 10^{-9}$~Hz. Nevertheless, the inclusion of binaries indeed introduces significant noise when $f \lesssim 10^{-7}\,{\rm Hz}$, corresponding to $P_b \gtrsim 24$~day. This noise grows as $P_b$ increases, coinciding with the behavior manifested by Eq.~(\ref{eq:Amp_binary}). Furthermore, there are clear breaks in the PSDs at $f = 2/T_{\rm obs}$, because the observed angular deflection of isolated stars or binary systems no longer exhibits periodic oscillations during the duration of Gaia's observations if $f \lesssim 2/T_{\rm obs}$, and consequently, two subtraction schemes will reduce the time-dependent  noise in the low-frequency region. The quadratic subtraction removes both the linear and quadratic components of the noise, resulting in lower noise PSDs compared to the linear subtraction. However, the quadratic subtraction also leads to more reductions to our signal at low frequencies, and we will see that the linear subtraction achieves better signal-to-noise ratio eventually.

\section{Astrometric deflections induced by GW and DPDM} \label{sec:deflection}
\subsection{GWs from binary systems}\label{subsec:GW}
To begin with, we revisit the formula for light deflection induced by GWs~(\cite{Book:2010pf, Pyne:1995iy}). 
The presence of GWs triggers metric perturbations, which lead to slight deflection of light traveling from a star to the Earth, ultimately resulting in the apparent displacement of stellar positions. By repeatedly measuring the apparent positions of a large number of stars across the sky, Gaia might be capable of capturing such deflection and thus extracting the characteristic patterns of GWs. 

The variation in the line of sight direction $n^i$ due to GWs can be expressed as~(\cite{Pyne:1995iy,Book:2010pf})
\begin{equation}
\delta n^i_{\rm GW}\simeq\frac{n^i_{} - p^i_{}}{2(1-\bm{n} \cdot \bm{p})} h^{\rm GW}_{j k}({\rm E}) n^j_{} n^k_{}-\frac{1}{2} h^{\rm GW}_{i j}({\rm E}) n^j_{} \; ,
\label{eq:deflection}
\end{equation}
where the unit vector $\bm{p}$ described by two rotational angles $\theta_p$ and $\varphi_p$ represents the propagated direction with $p^i$ being its components, and $h^{\rm GW}_{ij}(\rm E)$ refers to the metric perturbation calculated at the Earth. The metric perturbation $h^{\rm GW}_{ij}$ caused by monochromatic GWs with frequency $f_{\rm GW}$ reads
\begin{equation}
    h^{\rm GW}_{ij}(t,\bx) = {\rm Re}\left[ \mathcal{H}^{\rm GW}_{ij} e^{2\pi i f_{\rm GW} (t- {\bm p} \cdot \bm{x})}_{} \right] \; ,
    \label{eq:metric}
\end{equation}
with $\mathcal{H}^{\rm GW}_{ij} $ denoting the amplitude tensor. The GW possesses two polarization directions, and thus $\mathcal{H}^{\rm GW}_{ij}$ can be written as the following form
\begin{equation}
\mathcal{H}^{\rm GW}_{ij} = h^{\rm GW}_{+} \epsilon^{+}_{ij}(\bp, \psi) e^{i\phi^{}_+} + h^{\rm GW}_{\times}\epsilon^{\times}_{ij}(\bp,\psi)e^{i\phi^{}_\times} \; ,
\label{eq:tensor}
\end{equation}
where $\epsilon^{+}_{ij}$ and $\epsilon^{\times}_{ij}$ are two polarization tensors depending on $\bp$ and the GW polarization angle $\psi$~(\cite{NANOGrav:2023bts}), $h_{+}^{\rm GW}$ and $h_{\times}^{\rm GW}$  are their amplitudes, and $\phi^{}_+$ and $\phi_\times^{}$ are two corresponding phases.
We consider the monochromatic GW radiated from a binary system with a circular orbit, the amplitudes $h_+^{\rm GW}$ and $h_\times^{\rm GW}$ of which can be respectively expressed as
\begin{equation}
    h_+^{\rm GW} = h^{\rm GW}_0 \frac{1 + \cos^2 \iota}{2} \; , \quad 
    h_\times^{\rm GW} =  h^{\rm GW}_0 \cos \iota \; ,
    \label{eq:binary-ampli}
\end{equation}
where $h^{\rm GW}_0$ is nearly a constant depending on the chirp mass, the  angular orbital frequency and the luminosity distance to the source, and $\iota$ is the inclination angle of the binary. Finally, for the circular orbit, the relation $\varphi^{}_+ - \varphi^{}_\times = \pi/2$ should be satisfied. 

We omitted the ``star term" in Eq.~(\ref{eq:deflection}). According to \cite{Moore:2017ity}, the ratio between the ``star term" and the ``Earth term" equals the ratio between the wavelength of the GW and the distance to the source. For a GW frequency of $>10^{-10}\,{\rm Hz}$, the wavelength is $<100\, {\rm pc}$, making the ``star term" significantly smaller than the ``Earth term" assuming that the star is located at $1$ kpc away from us. Moreover, the ``star terms" from different sources are uncorrelated, resembling a kind of random noise, causing the ``star terms" being further suppressed after data compression, compared to the ``Earth term" that is consistently coherent.


\subsection{DPDM with purely gravitational interactions}\label{subsec:DP}
Analogous to GWs, dark photon fields can also induce metric fluctuations via gravitational effects~(\cite{Nelson:2011sf, Khmelnitsky:2013lxt, Nomura:2019cvc}). In this subsection, we derive the angular deflection of light emitted from the star as it passes through DPDM with purely gravitational interactions due to metric perturbations. 
We should mention that DPDM with a mass below $10^{-22}$~eV is subject to severe constraints from, e.g., the integrated Sachs-Wolfe effect on CMB anisotropies ($\gtrsim 10^{-24}~{\rm eV}$)~(\cite{Hlozek:2014lca}), the stellar kinematics of dwarf spheroidal galaxies (a few times of $10^{-23}$~eV)~(\cite{Gonzalez-Morales:2016yaf,Kendall:2019fep}) and the Lyman-$\alpha$ forest ($\gtrsim 10^{-21}~{\rm eV}$)~(\cite{Hui:2016ltb,Armengaud:2017nkf}). The lower bound on ultralight dark matter mass is still under debate, here we treat the local dark matter density $\rho_{\rm DM}$ as a free parameter.

We first describe the DPDM model. The dark photon field is expressed as
\begin{equation}
    {\bm A}(t,{\bm x}) \simeq A_0(\bx){\rm Re}\left[
    \bq(\bx)\exp( i m_A t)
    \right] \; ,
    \label{eq:DPDMfield}
\end{equation}
where the polarization vector $\bq(\bx) = (\cos\theta_q \cos\varphi_q e^{i \phi_1}$, $\cos\theta_q \sin\varphi_q e^{i \phi_2}, \sin\theta_q e^{i \phi_3})^{\rm T}$ is parameterized by two polarization angles $\theta_q$ and $\varphi_q$ and three phases $\phi_1$, $\phi_2$ and $\phi_3$, all of which are in general certain functions of $\bx$.
Under the synchronous gauge, the metric perturbation has the following form
\begin{equation}
\begin{aligned}
h_{ij}^{\rm DP}(t,\bx)\simeq {\rm Re}\left[ \mathcal{H}_{ij}^{\rm DP}(\bx)e^{2i m_A t}\right] ,
\label{eq:dpdm-metric}
\end{aligned}
\end{equation}
where
\begin{equation}
    \begin{aligned}
        \mathcal{H}_{ij}^{\rm DP}(\bx) = -h^{\rm DP}_0 \left[q_{i}(\bx)q_{j}(\bx)-\frac{\bq(\bx)\cdot\bq(\bx) \delta_{ij}}{4}\right].
    \end{aligned}
\end{equation}
The signal amplitude can be written as $h^{\rm DP}_0 = 4\pi G A^{2}_0(\bx)$, with $A_0(\bx)$ being  the amplitude of dark photon field. The amplitude of DPDM is
\begin{align}
    A_0(\bx) = \frac{\sqrt{2\rho_{\rm DM}(\bx)}}{m_A} \; ,
\end{align}
where $\rho_{\rm DM}(\bx)$ is the energy density of the DPDM. 
A detailed derivation of Eq.~(\ref{eq:dpdm-metric}) can be found in appendix~\ref{sec:appA}. As a result, we can estimate the signal amplitude
\begin{eqnarray}
    h^{\rm DP}_0 = 5.2\times 10^{-15}\left( \frac{\rho_{\rm DM}}{0.4 \, {\rm GeV}\,{\rm cm}^{-3}} \right)\left( \frac{ 10^{-23}\,{\rm eV}}{m_A}\right)^{2}.
\end{eqnarray}

It is known that the metric perturbation $h_{ij}^{\rm DP}$ given in Eq.~(\ref{eq:dpdm-metric}) yields a delay of the arrival time of the pulsar light, which can be probed by PTAs~(\cite{Nomura:2019cvc, PPTA:2021uzb, PPTA:2022eul, Xia:2023hov}). Here we show that the same metric perturbation could also lead to the deflection of star’s proper position, namely,
\begin{align}
    \delta n^i_{\rm DP} \simeq -\frac{(\delta^{ik}-n^in^k)}{2}n^j h_{jk}^{\rm DP}({\rm E}) + \mathcal{O}(v_{\rm vir}) \; ,
\end{align}
where we have omitted terms relying on the virial velocity $v_{\rm vir}$. The complete gauge-invariant expressions of $\delta n^i_{\rm DP}$ can be found in appendix~\ref{sec:appB}. The perturbation $h_{jk}^{\rm DP}({\rm E})$ is again calculated at the Earth and we ignored all the ``star terms". In Secs.~\ref{sec:case1}, \ref{sec:case2} and \ref{sec:case3}, we demonstrate that this approximation is valid as long as $m_A\geq 10^{-24}$\, eV. 

\subsection{DPDM with additional gauge couplings}\label{subsec:U1}
DPDM can also interact with ordinary matter if it is associated with ${\rm U(1)}^{}_{B}$ or ${\rm U(1)}^{}_{B-L}$ gauge symmetry. As a consequence, any baryonic matter could experience an oscillating force due to the DPDM background, leading to an acceleration $\ba$ given by~(\cite{Pierce:2018xmy})

\begin{equation}
    \bm{a} (t, \bm{x} ) \simeq \epsilon e \frac{q^{}_{\rm obj}}{m_{\rm obj}} \frac{\partial}{\partial t}  \bm{A}(t,\bx)\; ,
    \label{eq:accel}
\end{equation}
where $\epsilon$ is the coupling constant of the dark photon, $e$ denotes the electromagnetic coupling constant, and $q^{}_{\rm obj}$ refers to $B$ or $B-L$ charge the object carries, and $m^{}_{\rm obj}$ is the mass of the test object. 

The above acceleration leads to a velocity variation of the object
\begin{equation}
     \delta {\bm v}(t, {\bm x}) = \epsilon e \frac{q_{\rm obj}}{m_{\rm obj}} \bm{A}(t,\bx) \; ,
    \label{eq:velocity}
\end{equation}
which can cause aberrations of the stellar proper motion due to the moving observer~(\cite{Guo:2019qgs}). The resulting apparent angular deflection can be expressed as 
\begin{equation}
    \delta \bn_{B/B-L}(t,\bn) =  \delta\bvv(t,\bx_{\rm E}) - (\delta \bvv(t,\bx_{\rm E})\cdot\bn)\bn \; .\label{eq:angu-defl}
\end{equation}
where $\bx_{\rm E}$ is the coordinate of the observer, $\bn$ is the sky location of the star. Moreover, the dark photon field can also directly lead to stellar motion, with an amplitude approximated by $\delta \bn^s_{B/B-L} \lesssim |\delta \bvv| \cdot t / D$ (with $D$ being the distance between the star and the Earth). However, for the typical observation time $(\sim 1~{\rm year})$ and distance $D \gtrsim 1~{\rm kpc}$, $\delta\bn^s_{B/B-L} \sim 3 \times 10^{-4}\delta {\bm n}_{B/B-L}$ is rather tiny and will not be taken into consideration~(\cite{Guo:2019qgs}). 

\section{Sensitivity of GW and DPDM searches with Gaia} \label{sec:numerical}
In this section, we estimate the sensitivity of Gaia-like astrometry for GW and DPDM induced signals. We consider both Gaia and its next-generation upgrade (``XG-Gaia"), where we anticipate that the XG-Gaia will achieve $\mu$as angular resolution while maintaining the same number of observed stars as Gaia does (\cite{vallenari2018future}). The operational parameters of these missions are set as follows:
\begin{equation}
\begin{split}
    {\rm Gaia}:& \quad \sigma_{\rm ss} = 100~\mu{\rm as}\;, ~ \Delta t = 24~{\rm day}\;, \\
    & \quad T_{\rm obs}=10~{\rm year} \; , N =  1 \times 10^9 \; ;\\
    \text{XG-Gaia}:& \quad \sigma_{\rm ss} = 1~\mu{\rm as}\;, ~ \Delta t = 24~{\rm day}\;, \\
    & \quad T_{\rm obs}=20~{\rm year} \; , N =  1 \times 10^9 \; ,
    \nonumber
\end{split}
\end{equation}
where $\sigma_{\rm ss}$ is the noise from a single exposure,\footnote{The uncertainties on the proper motion parameters depend on stellar magnitudes $G$. It is expected the final uncertainty $\sigma_{\rm ss} \simeq 100~\mu{\rm as}$ for stars with $G \simeq 20$ in Gaia DR5~(\cite{Vallenari:2866069}). As these stars constitute the vast majority of the Gaia dataset, we will use $\sigma_{\rm ss} = 100~\mu{\rm as}$ for all stars later on.} $\Delta t$ is the observational cadence, $T_{\rm obs}$ is the total observation time, and $N$ is the total number of stars considered in the analysis.

As previously mentioned, we compress the data of $10^5$ stars in the same sky area by randomly generating $\widetilde{N}=10^4$ virtual stars uniformly distributed across the sky. Consequently, the instrument error for the measurement on each virtual star is rescaled to $\widetilde{\sigma}=\sigma_{\rm ss}/\sqrt{10^9/10^4}$. To estimate the sensitivity in detecting GWs and DPDM with Gaia-like experiments, we construct a mock dataset by injecting a specific GW or DPDM signal into the virtual stars. In the end of Sec.~\ref{subsec:GW}, Sec.~\ref{subsec:DP} and Sec.~\ref{subsec:U1}, we demonstrated that the signals we considered are consistently dominated by the ``Earth term" for $f_{\rm GW}>10^{-10}$ Hz or $m_A>10^{-24}$ eV.
Therefore, the astrometric effects on the stars in one area, caused by GWs or DPDM, can be approximated to the global effect on the virtual stars within that area.

In addition to the signal component, in Sec.~\ref{sec:background} we have established that the background motion due to stellar proper motion and the acceleration of the SSB is effectively removed by linear subtraction. The constant acceleration resulting from the galaxy's mass distribution is sufficiently small, thus can be safely ignored. Since linear subtraction is consistently applied to the astrometric data of virtual stars, background motion is not injected into the analysis. To address the time-dependent noise caused by stellar binaries, we adopt two different strategies for Gaia and XG-Gaia:

\begin{figure*}[t!]
    \centering		
    \includegraphics[width=\textwidth]{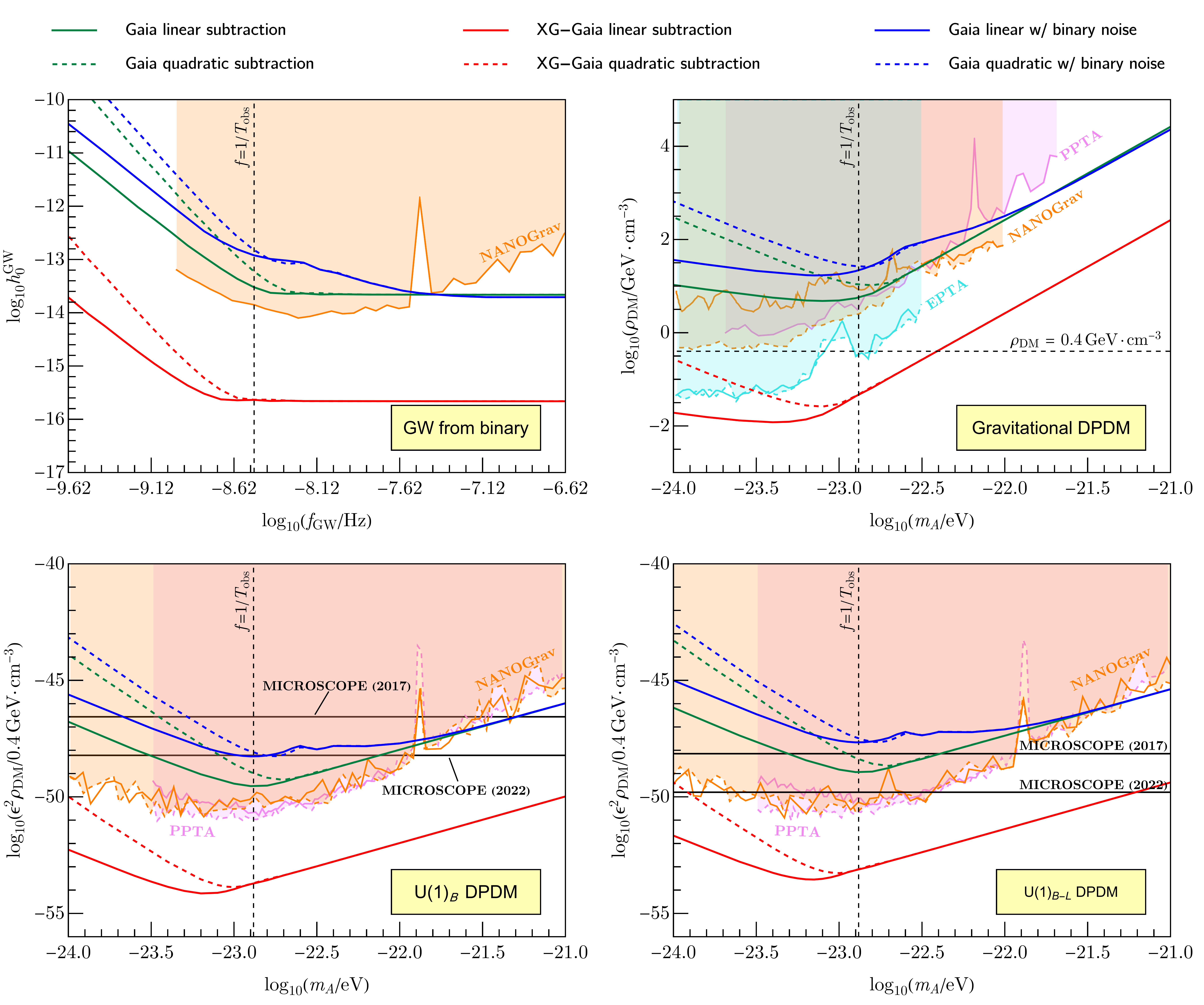} 	
    \vspace{0.cm}
    \caption{The 95\% exclusion limits from Gaia (green lines) and XG-Gaia (red lines) on the GW strain $h_0^{\rm GW}$ from a binary system, the energy density $\rho_{\rm DM}$ of the DPDM with purely gravitational interactions, and the coupling constant $\epsilon$ of $U(1)_{B}/U(1)_{B-L}$ DPDM, respectively. Model parameters regarding all the angles have been integrated out.
    Solid lines represent the bounds with linear subtraction, while dashed lines correspond to quadratic subtraction. We also use blue lines to denote the results including the time-dependent noise from unresolved binary systems. 
    For comparison, constraints from NANOGrav 15-year~(\cite{NANOGrav:2023pdq, NANOGrav:2023hvm}), EPTA DR2~(\cite{EuropeanPulsarTimingArray:2023egv}), and PPTA DR2~(\cite{PPTA:2021uzb, PPTA:2022eul}) results are labeled by orange, cyan and purple curves, respectively, with the solid (dashed) curves being the (un)correlated constraints. In the top-right panel, the black dashed line shows the amount of local dark matter density near the Earth as a reference. In the bottom two panels, the horizontal solid lines are the limits from the MICROSCOPE WEP experiment~(\cite{Touboul:2017grn,MICROSCOPE:2022doy, Amaral:2024tjg,Fayet:2017pdp,Fayet:2018cjy}). In addition, the vertical dashed line in each plot refers to the reference frequency $f = 1/T_{\rm obs}$ with $T_{\rm obs} = 10~{\rm year}$ being Gaia's observation period.}
    \label{fig:summary} 
    \vspace{0. cm}
\end{figure*}

\begin{itemize}[leftmargin=*]
\item For Gaia, we assume that all stellar binaries are unresolvable. Thus we use the PSD $S_{\rm bin}(f)$ calculated in Sec.~\ref{sec:background}, as shown in Fig.~\ref{fig:psd}.
\item For XG-Gaia, with $\sigma_{\rm ss} = 1~\mu{\rm as}$, binaries with net amplitudes $A_b \lesssim \sigma_{\rm ss}$ constitute only 0.01\% according to Eqs.~(\ref{eq:Amp_binary}) and (\ref{eq:Pb_distribution}). Therefore, we assume that almost all binaries have been resolved and removed from the data. Consequently, only single stars ($\sim$55\% of the total stars) are used for numerical simulation, and $S_{\rm bin}(f)=0$.
\end{itemize}

We construct $\chi^2$-functions for both the signal hypothesis and null hypothesis in the frequency domain, namely,
\begin{equation}
    \chi^2_{s} =  \frac{2}{T_{\rm obs}}\sum_{I,K} \frac{|\widetilde{\bm d}_{I}(f_K)-\widetilde{\bm s}_I(\Phi;f_K)|^2}{ S_{n}(f_K)} \; , 
    \label{eq:chi-signal}
\end{equation}
and
\begin{equation}
    \chi^2_{n} =  \frac{2}{T_{\rm obs}}\sum_{I,K} \frac{|\widetilde{\bm d}_{I}(f_K)|^2}{ S_{n}(f_K)} \; ,
    \label{eq:chi-noise}
\end{equation}
where $\widetilde{\bm d}_I$ is the astrometry data of the virtual star $I$ after linear or quadratic subtraction, $\widetilde{\bm s}_I (\Phi;f_K)$ represents the model predictions for a given set of parameters $\Phi$ after the same subtraction strategy, and $S_{n}(f)$ is defined in Eq.~(\ref{eq:psd_n}) with $S_{w}(f) = \widetilde{\sigma}\Delta t$. Note that both $\widetilde{\bm d}_I$ and $\widetilde{\bm s}_I$ are the quantities after Fourier transforms, and we sum over $I$ and $K$, which correspond to the number of virtual stars and that of bins in the PSD, respectively.  The difference between the signal hypothesis and null hypothesis, i.e., $\Delta \chi^2 = \chi^2_{s} -\chi^2_{n}$ becomes an indicator for measuring the significance of the signal.

The free parameters in $\Phi$ for (A) GWs from binary systems, (B) DPDM with purely gravitational interactions, and (C) DPDM with ${\rm U}(1)_{B}$ or ${\rm U}(1)_{B-L}$ gauge symmetry are respectively 
\begin{equation}
    \begin{split}
    \text{(A)}:&  \quad \{ f_{\rm GW}, \; h_0^{\rm GW}, \; \theta_p \;, \varphi_p, \; \iota, \;\psi, \; \phi_+ \}\;; \\
    \text{(B)}:& \quad \{ m_{A}, \; \hat{\rho}_{\rm DM},  \; \theta_q, \; \varphi_q, \; \phi_1 \; , \phi_2 \; , \phi_3\} \; ; \\
    \text{(C)}:& \quad \left\{ m_{A}, \; \epsilon^2 \hat{\rho}_{\rm DM}, \; \theta_q, \; \varphi_q, \; \phi_1 \; , \phi_2 \; , \phi_3 \right\} \; ,
    \nonumber
\end{split}
\end{equation}
where 
\begin{equation}
    \hat{\rho}_{\rm DM}\equiv\frac{\rho_{\rm DM}}{0.4 \,{\rm GeV}\,{\rm cm}^{-3}}.
\end{equation}
We fix the values of $f_{\rm GW}$ and $m_A$, and scan the other model parameters in their individual parameter space in each case. The
95\% exclusion limits on the $h_0^{\rm GW}$, $\rho_{\rm DM}$ and $\epsilon$ after integrating out all angular parameters are shown in Fig.~\ref{fig:summary}. Some remarks are in order. Firstly, the frequency $f = 2/T_{\rm obs}$  approximately marks where the sensitivity curves show significant changes. In the high-frequency range ($f \gtrsim 2/T_{\rm obs}$), the limits obtained using two different subtraction methods are indistinguishable, as expected, whereas linear subtraction offers better sensitivity in the low-frequency range compared to quadratic subtraction. In particular, the 95\% exclusion limits are enhanced by more than one order of magnitude at $f_{\rm GW} \lesssim 6 \times 10^{-10}\,{\rm Hz}$ for GWs, or at $m_A \lesssim 3\times 10^{-24}\,{\rm eV}$ for DPDM. 
Secondly, the impact of time-dependent noise from unresolved binary systems significantly decreases the sensitivity by up to an order of magnitude. Hence the inclusion of such noise is important in our analysis. In addition, this decrease is more severe in the linear subtraction scheme than in the quadratic subtraction, as we explained in Sec.~\ref{sec:background}. 
Lastly, XG-Gaia will possess extended observation duration and improved precision. It will also have enhanced resolution capabilities for binary systems, which is crucial for suppressing the noise from unresolved binaries. As a result, the sensitivity of XG-Gaia to detecting GWs and DPDM is expected to increase by more than three orders of magnitude.

\section{Conclusion} \label{sec:conclusion}

Astrometric measurements offer a new approach to detecting signals such as GWs and DPDM in the nHz frequency range. Typically, periodic signals with frequencies below $1/T_{\rm obs}$ are easily mistaken for background noise or systematic uncertainties. Known background sources include (1) the proper motion of stars, (2) stellar acceleration due to the Milky Way’s mass distribution, (3) acceleration from companion stars in binary systems, and (4) the Earth's acceleration. In Sec.~\ref{sec:background}, we demonstrate that a linear subtraction method is sufficient to remove all these effects. Compared to the commonly used quadratic subtraction, linear subtraction significantly enhances detection sensitivity in the low-frequency range.

However, other factors, such as telescope movement, changes in satellite orientation, and unknown systematics, are not investigated here. We project detection prospects in this study, assuming all these systematics are properly subtracted. We emphasize that future detection claims will require more detailed studies of these systematics. Furthermore, this work does not explore the angular patterns induced by gravitational waves and dark photon dark matter, which may be crucial for distinguishing between systematics and genuine signals.


Astrometric data can also be correlated with the current existing PTA data, allowing for a more concrete validation of the correlation pattern induced by GWs or dark matter and a more sensitive probe of macroscopic polarizations of the GWs (\cite{Qin:2018yhy,Qin:2020hfy,Caliskan:2023cqm}). Compared to PTAs, future astrometric surveys like Roman (\cite{Wang:2020pmf,Wang:2022sxn}) is able to probe $\mu$Hz GWs or beyond and ultralight bosons with higher masses, allowing for cross-correlation with other proposed measurements in the same frequency range such as binary resonance (\cite{Blas:2021mpc,Blas:2021mqw}) and Doppler tracking (\cite{Zwick:2024hag}), potentially closing the gap between PTA and LISA frequency ranges.

\section*{Acknowledgement}
We would like to thank Huai-Ke Guo for his collaboration in the early stages of this project. We also thank Christopher J. Moore for useful email exchanges. H.A. is supported in part by the National
Key R\&D Program of China under Grant No.
2023YFA1607104 and 2021YFC2203100, the National Natural
Science Foundation of China (NSFC) under Grant No.
11975134, and the Tsinghua University Dushi Program.  J.S. is supported by Peking University under startup Grant No. 7101302974 and the National
Natural Science Foundation of China under Grants No.
12025507, No.~12150015; and is supported by the Key
Research Program of Frontier Science of the Chinese
Academy of Sciences (CAS) under Grants No. ZDBS-LY7003. X.W. acknowledges the Royal Society as the
funding source of the Newton International Fellowship.  X.X. is supported by Deutsche
Forschungsgemeinschaft under Germany’s Excellence Strategy EXC2121 “Quantum Universe”
— 390833306. Y.Z. is supported by the U.S. Department of Energy under Award No.
DESC0009959.

\vspace{0.5 cm}

{\bf Note added}: During the preparation of this paper, we became aware of the recent work by (\cite{Kim:2024xcr}) and (\cite{Dror:2024con}) which considered the astrometric effect of the ultralight scalar dark matter. Our work focuses on the detection of the ultralight vector dark matter and continuous gravitational waves, which is distinctive from their themes. 

\clearpage
\appendix
\widetext

\section{Metric perturbation induced by DPDM} \label{sec:appA}
Consider a general metric perturbation $h_{\mu\nu}$, the equation of motion can be written as (\cite{Carroll:2004st})
\begin{align}
    \Box h_{\mu\nu} - \partial^{\alpha}\partial_{\nu}h_{\mu\alpha} 
    - \partial^{\alpha}\partial_{\mu}h_{\nu\alpha} 
    +\partial_{\mu}\partial_{\nu}h
    + \eta_{\mu\nu}(\partial^{\alpha}\partial^{\beta}h_{\alpha\beta}
    -\Box h) = -16\pi G T_{\mu\nu} \; ,
\end{align}
where $h\equiv h_{\mu\nu}\eta^{\mu\nu}$ and $\Box \equiv \partial^{\mu}\partial_{\mu}$. We work in the Minkowski background $\eta_{\mu\nu}={\rm diag}(-1,1,1,1)$. Under the gauge fixing condition $h_{0\mu}=0$, the Einstein equations become
\begin{align}
       -16\pi G T_{00}& = \partial_0^2 h -(\partial_k \partial_l h_{kl}-\Box h) \; ,\label{eq1}\\
      -16\pi G T_{0i}&=-\partial_k\partial_0 h_{ik}+\partial_i\partial_0 h \; ,\label{eq2}\\
     -16\pi GT_{ij}&=\Box h_{ij} - \partial_k\partial_ih_{jk}- \partial_k\partial_jh_{ik}+ \partial_i\partial_jh + \delta_{ij}\partial_k\partial_lh_{kl}- \delta_{ij}\Box h \; .\label{eq3}
    \end{align}
From Eq.~(\ref{eq3}) we find
\begin{align}
    -2\Box h +\nabla^2 h + \partial_l\partial_jh_{jl} = -16\pi G T_{kk}\;.\label{eq4}
\end{align}
It is also convenient to derive
\begin{equation}
    -16\pi G\left(T_{ij}-\frac{\delta_{ij}}{2}T_{kk}\right) =\Box h_{ij} - \partial_k\partial_ih_{jk}  - \partial_k\partial_jh_{ik}+ \partial_i\partial_jh  
     + \frac{1}{2}\delta_{ij}\partial_k\partial_lh_{kl}
    -\frac{1}{2}\delta_{ij}  \nabla^2 h \; .\label{eq5}
\end{equation}

Next we calculate the stress-energy tensor induced by the DPDM. 
The virial velocity of our galaxy is $v_{\rm vir}\simeq {\cal O}(10^{-3})$, so for ultralight 
DPDM with mass $m_A \sim \mathrm{10^{-23}~eV}$, the de Broglie wavelength is $\lambda_{dB}\sim 4 \, \mathrm{kpc}\,(10^{-23}\,\mathrm{eV}/m_A)(10^{-3}/v_{\rm vir})$, which is of the same order as the galactic size. The coherence time is $t_c\sim 1/(m_A v^2_{\rm vir})=2 \, \mathrm{Myr}\,(10^{-23}\,\mathrm{eV}/{m_A})(10^{-3}/v_{\rm vir})^2$. Since we are considering DPDM within a coherent time, we can write down the spatial component of dark photon field as
\begin{align}
    &\bm{A}(t,\bx) = A_0(\bx){\rm Re}\left[\bq
    (\bx)e^{iEt}
    \right] \; ,
\end{align}
where $E = m_A\sqrt{1 + \mathcal{O}(v_{\rm vir}^2)}$ , $\bq(\bx)$ is a complex vector that characterizes the polarization of the dark photon field, normalized as $|\bq|=1$. The spatial derivative of $A(\bx)$ and $\bq(\bx)$ is suppressed by virial velocity $v_{\rm vir}$. Using the equation of motion $\partial_\mu A^{\mu}=0$, we find that the scalar potential of the dark photon field $A^t$ is suppressed by $v_{\rm vir}$
\begin{align}
    A^t =  \mathcal{O}(v_{\rm vir}) \; .
\end{align}
The stress-energy tensor of vector field reads
\begin{align}
    T_{\mu\nu}(t,\bx)=\eta_{\mu\nu}\left(-\frac{1}{4}F_{\rho\sigma}F^{\rho\sigma}-\frac{1}{2}m^{2}A^{\rho}A_{\rho}\right)
+\eta^{\rho\sigma}F_{\mu\sigma}F_{\nu\rho}+m^{2}A_{\mu}A_{\nu} \; .
\end{align}
We find that at the zeroth order of $v_{\rm vir}$, the relevant components of the stress-energy tensor are
\begin{equation}
\begin{aligned}
    &T_{00}^{(0)} = \rho(\bx) \; ,\qquad T_{0k}^{(0)} = 0 \;,\\
    &T^{(0)}_{kk}= -\rho(\bx){\rm Re}\left[
    \bq(\bx) \cdot \bq(\bx)   \,e^{2im_At}
    \right] \; ,\\
    &T_{ij}^{(0)}-\frac{\delta_{ij}}{2}T_{kk} = 2\rho(\bx){\rm Re}\left\{
    \left[q_i(\bx) q_j(\bx) -\frac{\bq(\bx) \cdot\bq(\bx)  \delta_{ij}}{4}\right]e^{2im_At}
    \right\} \; . \label{eqtij}
\end{aligned}
\end{equation}
 where for simplicity we defined $\rho(\bx)\equiv m^2_A A_0^2(\bx)/2$. By combining Eq. (\ref{eq1})  and Eq. (\ref{eq4}), we can eliminate $\partial_l \partial_k h_{kl}$ and get (hereafter we add superscript ``DP" for quantities associated with the dark photon dark matter)
\begin{equation}
\partial_0^2 h^{\rm DP} = -8\pi G \left(T_{00}+T_{kk}\right) 
=-8\pi G \rho(\bx)\left\{1-{\rm Re}\left[
    \bq(\bx) \cdot \bq(\bx)  \, e^{2im_At}
    \right]\right\} \; .\label{trace}
\end{equation}
Integrating Eq. (\ref{trace}), we find that the trace $h$ contains at least an oscillating component
\begin{equation}
\begin{aligned}
h^{\rm DP}(\bx) = -\frac{8\pi G \rho(\bx)}{m^2_A}{\rm Re}\left[
   \frac{\bq(\bx)\cdot \bq(\bx)}{4}  \,e^{2im_At}
   \right] \; ,\label{trace-1}
\end{aligned}
\end{equation}
which suggests that $h_{ij}$ also has an oscillating part and a static part
\begin{align}
    h_{ij}^{\rm DP}(t,\bx)=-h_0^{\rm DP}(\bx){\rm Re}\left[\mathcal{H}^{\rm DP}_{ij}(\bx)
      \,e^{2im_At}
    \right] \; .\label{eq:hij_general}
\end{align}
Combining Eq. (\ref{eq5}) and Eq. (\ref{eqtij}), we obtain
\begin{align}
-32\pi G \rho(\bx){\rm Re}\left\{
    \left[q_i(\bx)q_j(\bx)-\frac{\bq(\bx)\cdot\bq(\bx) \delta_{ij}}{4}\right]\,e^{2im_At}
    \right\} =\Box h_{ij}^{\rm DP} &- \partial_k\partial_ih_{jk}^{\rm DP}- \partial_k\partial_jh_{ik}^{\rm DP}+ \partial_i\partial_jh^{\rm DP} \nonumber \\ &+ \frac{1}{2}\delta_{ij}\partial_k\partial_lh_{kl}^{\rm DP}
    -\frac{1}{2}\delta_{ij}  \nabla^2 h^{\rm DP} \; , \label{eq7}
\end{align}
 which is accurate to the zeroth order of $v_{\rm vir}$. We extract the oscillation parts on both sides of Eq. (\ref{eq7}), which yields
 \begin{align}
\partial_0^2 {\rm Re}\left[-h_0^{\rm DP}(\bx)\mathcal{H}^{\rm DP}_{ij}(\bx)
     \,e^{2im_At}
    \right] = 32\pi G \rho(\bx){\rm Re}\left\{
    \left[q_i(\bx)q_j(\bx)-\frac{\bq(\bx)\cdot\bq(\bx) \delta_{ij}}{4}\right]\,e^{2im_At}
    \right\} \; ,
\end{align}
 where we neglect the terms with spatial derivatives since they are suppressed by $v_{\rm vir}$. Solving this equation, we get
\begin{equation}
    \begin{aligned}
    \mathcal{H}^{\rm DP}_{ij}(\bx) = q_i(\bx)q_j(\bx)-\frac{\bq(\bx)\cdot\bq(\bx) \delta_{ij}}{4} \; .
    \end{aligned}
\end{equation}
and
\begin{equation}
    \begin{aligned}
        h_0^{\rm DP}(\bx)=\frac{8\pi G \rho(\bx)}{m^2_A}\;.
    \end{aligned}
\end{equation}

\section{Gauge invariance of the angular deflection}\label{sec:appB}

In this appendix we calculate the angular deflection of light from stars due to the oscillating metric perturbations. This part follows the discussion in~\cite{Book:2010pf}. In this part, we also prove the gauge invariance of the angular deflection. Consider a general metric perturbation
\begin{align}
    \mathrm{d}s^{2}=g_{\mu\nu}\mathrm{d}x^{\mu}\mathrm{d}x^{\nu}=(\eta_{\mu\nu}+h_{\mu\nu})\mathrm{d}x^{\mu}\mathrm{d}x^{\nu} \; .
\end{align}
where we work in the Minkowski background with $\eta_{\mu\nu}={\rm diag}\{ -1,1,1,1\}$. Now consider a photon travels from the source to the detector, then the angular deflection of light we need to calculate becomes angular difference of the observed photon momentum compared to the flat spacetime case. Without metric perturbation, we choose the detector to be at the origin $(0, \boldsymbol{0})$ and the unperturbed photon worldline to be $x_{(0)}^{\mu}=\omega_{0}\mathcal{N}^{\mu}\lambda$, where $\bn$ is the unit vector of the source, $\omega_0$ is the unperturbed photon frequency, and $\mathcal{N}^{\mu}=(1,-\bn)$. The unperturbed photon four-momentum is $k_{(0)}^{\mu}=\omega_{0}\mathcal{N}^{\mu}$. Suppose the unperturbed position of the source is  $x_s^i(t) = x_s^i$, then the affine parameter of the source is $\lambda_s=-|\bx_s|/\omega_0$.

We separate the photon worldline into unperturbed parts and the first-order perturbation
\begin{align}
    k^{\mu}(\lambda)=k_{(0)}^{\mu}(\lambda)+k_{(1)}^{\mu}(\lambda) + \mathcal{O}(h^2) \; ,\label{eq:four-momentum}
\end{align} To the first order of $h$, the geodesic equation of photon is
\begin{align}
    \frac{\mathrm{d}k^{\rho}}{\mathrm{d}\lambda}=-\Gamma^{\rho}_{\mu \nu}k^{\mu}_{(0)}k^{\nu}_{(0)} \; .
\end{align}
We integrate the above expression from $\lambda=0$ to arbitrary $\lambda$ to obtain $k^{\mu}(\lambda)$
\begin{align}
    k^{\mu}(\lambda)=k_{(0)}^{\mu} + \kappa^{\mu} -\int_{0}^{\lambda}\Gamma^{\rho}_{\mu \nu}k^{\mu}_{(0)}k^{\nu}_{(0)} \mathrm{d}\lambda^{\prime} \; ,\label{eq:four-momentum_expression}
\end{align}
where $\kappa^{\mu}$ is the first order perturbation of the photon four-momentum at $\lambda=0$. Integrating Eq. (\ref{eq:four-momentum_expression}) yields the trajectory of photon
\begin{align}
    x^{\mu}(\lambda)=\lambda (k_{(0)}^{\mu} + \kappa^{\mu}) -\int_{0}^{\lambda}\int_{0}^{\lambda^{\prime}}\Gamma^{\rho}_{\mu \nu}(\lambda^{\prime \prime})k^{\mu}_{(0)}k^{\nu}_{(0)} \mathrm{d}\lambda^{\prime \prime}\mathrm{d}\lambda^{\prime} + x^{\mu}(0) \; ,\label{eq:trajectory}
\end{align}
where $x^{\mu}(0)$ is the final position of the photon, which also refers to the position of the detector.
Choosing $\lambda=\lambda_{s}$ in Eq. (\ref{eq:trajectory}), we are able to solve $\kappa^{\mu}$
\begin{align}
    \kappa^{\mu}=\frac{1}{\lambda_s}\left(x^{\mu}(\lambda_s)-x^{\mu}(0)+\int_{0}^{\lambda_s}\int_{0}^{\lambda^{\prime}}\Gamma^{\mu}_{\gamma \nu}(\lambda^{\prime \prime})k^{\gamma}_{(0)}k^{\nu}_{(0)} \mathrm{d}\lambda^{\prime \prime}\mathrm{d}\lambda^{\prime}\right) - k_{(0)}^{\mu} \; . \label{eq:initial-k}
\end{align}
To the zeroth order of perturbation, $x^{\mu}(\lambda_s)-x^{\mu}(0)=\lambda_s k_{(0)}^{\mu}$. With the presence of metric perturbation, we can not assume that the detector and source are static since the geodesic equations of their spatial trajectories also contain $\mathcal{O}(h)$ perturbations.
Therefore we keep $x^{\mu}(\lambda_s)$ and $x^{\mu}(0)$ in the expression of $\kappa^{\mu}$.

In addition, the null geodesic equation constraint $g_{\mu\nu}(\lambda)k^{\mu}(\lambda) k^{\nu}(\lambda)=0$ provide
\begin{equation}
    \kappa_0 = n_i \kappa_i - \frac{1}{2\omega_0}h_{\mu\nu}(0)k^{\mu}_{(0)}k^{\nu}_{(0)} \; .
    \label{eq:null 2}
\end{equation}

To calculate the angular deflection the detector observes, we need to project the photon four-momentum into the local 
proper reference frame of the detector. We define the four-momentum of the detector as $e^{\nu}_0$ and a set of orthonormal basis vectors as $e^{\nu}_i, i=1,2,3$. We decompose the basis vectors into unperturbed parts and the first-order perturbations
\begin{align}
e^{\nu}_{\alpha}=\delta^{\nu}_{\alpha}+\delta e^{\nu}_{\alpha} \; , \quad \alpha=0,1,2,3 \; .\label{eq:e_perturbation}
\end{align}
The basis vectors of local proper reference frame satisfy
\begin{equation}
e^{\mu}_{0}\nabla_{\mu}e^{\nu}_{\alpha}=0 \; .
    \label{eq:basis-local}
\end{equation}
Inserting Eq.~(\ref{eq:e_perturbation}) into Eq. (\ref{eq:basis-local}), we find
\begin{align}
    \partial_{0}\delta e^{\nu}_{\alpha}+\Gamma_{\alpha 0 }^{\nu}=0 \; .
    \label{eq:e comdition}
\end{align}
The orthonormal conditions of $e^{\nu}_{\alpha}$ implies
\begin{align}
    g_{\mu\nu}e^{\nu}_{\alpha}e^{\mu}_{\beta}=\delta_{\alpha\beta} \; .
    \label{eq:basis-2}
\end{align}
We keep only the first order terms of Eq. (\ref{eq:basis-2}) which gives another constraint on $\delta e^{\nu}_{\alpha}$
\begin{align}
    h_{\mu\nu} + \eta_{\mu\gamma}\delta e^{\gamma}_{\nu}+\eta_{\nu\gamma}\delta e^{\nu}_{\mu}=0  \; .
    \label{eq:basis-c}
\end{align}

The angular deflection of the dark photon reads
\begin{align}
    \delta n_{i} = -\frac{g_{\nu\mu}(0) k^{\nu}(0) e^{\mu}_{i}}{\omega_{d}} -n_{i}\;,
    \label{eq:angle-1}
\end{align}
where $\omega_{d}=-g_{\nu\mu}(0) k^{\nu}$ is the observed photon frequency. Expanding Eq. (\ref{eq:angle-1}) to the first order, we obtain
\begin{align}
    \delta n_{i} = -\frac{1}{\omega_0}\left(h_{\nu i}(0)k^{\nu}_{(0)}+\kappa_i+\eta_{\nu\mu}k^{\nu}_{(0)}\delta e^{\mu}_{i}-n_i(h_{\nu 0}(0)k^{\nu}_{(0)}+\eta_{\nu\mu}k^{\nu}_{(0)}\delta e^{\mu}_{0}+\kappa_0)\right) \; .
    \label{eq:angle-2}
\end{align}
Plugging Eq. (\ref{eq:null 2}) into Eq.~(\ref{eq:angle-2}), we find 
\begin{align}
    \delta n_{i} = -\frac{1}{\omega_0}P_{ij}\kappa_{j}-h_{\nu i}(0)\mathcal{N}^{\nu}-\mathcal{N}_{\mu}\delta e^{\mu}_{i}+n_i\left(h_{\nu 0}(0)\mathcal{N}^{\nu}+\mathcal{N}_{\mu}k_{d(1)}^{\mu}\right)-\frac{1}{2}n_i h_{\nu\mu} \mathcal{N}^{\mu} \mathcal{N}^{\nu} \; ,
    \label{eq:angle-3}
\end{align}
where $P_{ij}=\delta_{ij}-n_i n_j$ is the projection operator. Multiplying both sides of Eq. (\ref{eq:angle-3}) by $n^i$, we obtain
\begin{align}
    n^i \delta n_{i} &= -n^i h_{\nu i}(0)\mathcal{N}^{\nu}-n^i \mathcal{N}_{\mu}\delta e^{\mu}_{i}+h_{\nu 0}(0)\mathcal{N}^{\nu}+\mathcal{N}_{\mu}\delta e_{0}^{\mu}-\frac{1}{2}h_{\nu\mu} \mathcal{N}^{\mu} \mathcal{N}^{\nu} \nonumber \\
            &=- n^i \mathcal{N}_{\mu}\delta e^{\mu}_{i}+\mathcal{N}_{\mu}\delta e_{0}^{\mu}+\frac{1}{2}h_{\nu\mu} \mathcal{N}^{\mu} \mathcal{N}^{\nu} \nonumber  \\
            &=\mathcal{N}_{\mu} \mathcal{N}^{\nu}\delta e^{\mu}_{\nu}+\frac{1}{2}h_{\nu\mu} \mathcal{N}^{\mu} \mathcal{N}^{\nu} \nonumber  \\
            &=0 \; .
    \label{eq:angle-o}
\end{align}
Note that in the last step we use Eq. (\ref{eq:basis-c}). Now $n^i \delta n_{i}=0$ shows that $\delta n_{i}$ is orthogonal to $n^i$, so we can multiply $P_{ij}$ to each term 
in the expression of $\delta n_{i}$, which reads
\begin{align}
    \delta n_{i} = -P_{ij}\left(\frac{1}{\omega_0}\kappa_{j}+h_{\nu j}(0)\mathcal{N}^{\nu}+\mathcal{N}_{\mu}\delta e^{\mu}_{j}\right) \; .
    \label{eq:angle-4}
\end{align}
Then we prove that this expression is gauge-invariant. Consider a transformation of coordinates $x^{\prime\mu}=x^{\mu}+\xi^{\mu}(x)$, it induces a gauge transformation of linearized theory
\begin{equation}
    h_{\mu\nu}(x)\rightarrow h_{\mu\nu}(x)-(\partial_{\mu}\xi_{\nu}(x)+\partial_{\nu}\xi_{\mu}(x))\; .
    \label{eq:gauge tr1}
\end{equation}

We check the gauge transformation of each term in $\delta n_i$. We use $\Delta$ to represent the change of each term after gauge transformation.
Notice that $\Delta \Gamma^{\rho}_{\mu\nu}=-\partial_{\mu}\partial_{\nu}\xi^{\rho}$, $\Delta x^{\mu}(\lambda)=\xi^{\mu}(\lambda)$, we have
\begin{align}
    \Delta \kappa^{\mu} &= \frac{1}{\lambda_s}\left(\xi^{\mu}(\lambda_s)-\xi^{\mu}(0)-\int_{0}^{\lambda_s}\int_{0}^{\lambda^{\prime}}\partial_{\gamma}\partial_{\nu}\xi^{\mu}(\lambda^{\prime \prime})k^{\gamma}_{(0)}k^{\nu}_{(0)} \mathrm{d}\lambda^{\prime \prime}\mathrm{d}\lambda^{\prime}\right) \nonumber \\
    &= k^{\nu}_{(0)}\partial_{\nu}\xi^{\mu}(0)=\omega_0 \mathcal{N}^{\nu}\partial_{\nu}\xi^{\mu}(0) \; .
    \label{eq:angle-5}
\end{align}
In the rest of the discussion we ignore the coordinate dependence of $\xi^{\mu}$ since now all the $\xi^{\mu}(\lambda)$ are defined on $\lambda=0$.
Here we use $k^{\nu}_{(0)}\partial_{\nu}\xi^{\mu}=\mathrm{d}\xi^{\mu}/\mathrm{d}\lambda$ to simplify the integration.

From Eq. (\ref{eq:e comdition}) we find
\begin{equation}
    \partial_{0}(\Delta \delta e^{\nu}_{i})=-\Delta \Gamma_{i 0 }^{\nu}=\partial_0 \partial_{i}\xi^{\nu} \; ,
    \label{eq:gauge tr2}
\end{equation}
which means $\Delta \delta e^{\nu}_{i}=\partial_{i}\xi^{\nu}+f$. $f$ is an arbitrary function that has no dependence on time. We ignore $f$ because it is not correlated with 
gauge transformation. 
Insert these expressions into $\delta n_{i}$, we can get the gauge transformation of the angular deflection.
\begin{align}
    \Delta\delta n_{i} = -P_{ij}\left(\mathcal{N}^{\nu}\partial_{\nu}\xi_{j}-(\partial_{j}\xi_{\nu}+\partial_{\nu}\xi_{j})\mathcal{N}^{\nu}+\mathcal{N}_{\mu}\partial_j \xi^{\mu}\right)=0 \; .
    \label{eq:angle-gauge}
\end{align}
Therefore the angular deflection $\delta n_{i}$ is gauge-invariant.
Now we choose a typical gauge to calculate $\delta n_i$. We use synchronous gauge which is defined by $h_{00}=h_{0i}=0$.
The geodesic equations of the four-momentum of the detector is reduced to 
\begin{equation}
    \partial_{0}\delta e^{\nu}_{\mu}=-\Gamma_{\mu 0 }^{\nu}=-\frac{1}{2}\partial_0 h_{\mu}^{\nu} \; ,
\end{equation}
which gives
\begin{equation}
    \delta e^{\nu}_{\mu}=-\frac{1}{2} h_{\mu}^{\nu} \; .
\end{equation}
For $\mu=0$, $\delta e^{\nu}_{0}=0$. Thus the spatial components of the four-momentum of the detector do not contain first order perturbations.
Then we can treat the detector as a static object and $x^{\mu}(0)$ becomes a constant. Similarly, $x^{\mu}(\lambda_s)$ is also a constant, and they 
satisfy $x^{\mu}(\lambda_s)-x^{\mu}(0)=\lambda_s k^{\mu}_{(0)}$. The final expression of angular deflection is
\begin{align}
    \delta n_{i} &= -P_{ij}\left(\frac{\eta_{j\mu}}{\omega_0 \lambda_s}\int_{0}^{\lambda_s}\int_{0}^{\lambda^{\prime}}\Gamma^{\mu}_{\gamma \nu}(\lambda^{\prime \prime})k^{\gamma}_{(0)}k^{\nu}_{(0)} \mathrm{d}\lambda^{\prime \prime}\mathrm{d}\lambda^{\prime}-\frac{1}{2}h_{ jk}(0)n^{k}\right) \nonumber \\
    &= -P_{ij}\left(\frac{\omega_0}{ \lambda_s}\int_{0}^{\lambda_s}\int_{0}^{\lambda^{\prime}}\left(-n^l\partial_0h_{jl}+n^l n^k\partial_k h_{jl}-\frac{1}{2}n^l n^k \partial_j h_{kl}\right) \mathrm{d}\lambda^{\prime \prime}\mathrm{d}\lambda^{\prime}-\frac{1}{2}h_{ jk}(0)n^{k}\right) \; .
    \label{eq:angle-syn}
\end{align}
Using the identity
\begin{align}
    \frac{\mathrm{d}}{\mathrm{d}\lambda}h_{jl}=\omega_0\left(\partial_0 h_{jl}-n^k\partial_k h_{jl}\right) \; ,
\end{align}
we can integrate the first two terms in Eq. (\ref{eq:angle-syn})
\begin{align}
    \delta n_{i} &= -P_{ij}\left(\frac{1}{ \lambda_s}\int_{0}^{\lambda_s}\left(-n^l h_{jl}(\lambda^{\prime})+n^lh_{jl}(0)\right) \mathrm{d}\lambda^{\prime}-\frac{\omega_0}{ 2\lambda_s}\int_{0}^{\lambda_s}\int_{0}^{\lambda^{\prime}}n^l n^k \partial_j h_{kl}\mathrm{d}\lambda^{\prime \prime}\mathrm{d}\lambda^{\prime}-\frac{1}{2}h_{ jk}(0)n^{k}\right) \nonumber \\
    &=-P_{ij}\left(-\frac{1}{ \lambda_s}\int_{0}^{\lambda_s}n^l h_{jl}(\lambda^{\prime}) \mathrm{d}\lambda^{\prime}-\frac{\omega_0}{ 2\lambda_s}\int_{0}^{\lambda_s}\int_{0}^{\lambda^{\prime}}n^l n^k \partial_j h_{kl}\mathrm{d}\lambda^{\prime \prime}\mathrm{d}\lambda^{\prime}+\frac{1}{2}h_{ jk}(0)n^{k}\right)\; .
    \label{eq:angle-syn1}
\end{align}

The angular deflection is separated into three terms
\begin{align}
    \delta n_{1i}&=\frac{1}{ \lambda_s}P_{ij}\int_{0}^{\lambda_s}n^l h_{jl}(\lambda^{\prime}) \mathrm{d}\lambda^{\prime}\;,\\
    \delta n_{2i}&=\frac{\omega_0}{ 2\lambda_s}P_{ij}\int_{0}^{\lambda_s}\int_{0}^{\lambda^{\prime}}n^l n^k \partial_j h_{kl}(\lambda^{\prime\prime})\mathrm{d}\lambda^{\prime \prime}\mathrm{d}\lambda^{\prime}\;,\\
    \delta n_{3i}&=-\frac{1}{2}P_{ij}h_{ jk}(0)n^{k}\; .\label{eq:three_dn}
\end{align}

Now we estimate the relative size of these three terms. Without loss of generality, we consider the metric perturbation that takes the following form
\begin{align}
    h_{ij}^{\rm DP}(\bx,t)=  c\mathcal{H}_{ij}^{\rm DP}\rho(\bx)\cos(2 m _At+2\alpha)\;,\label{eq:hij-express}
\end{align}
where $\mathcal{H}_{ij}^{\rm DP}$ is an order one tensor,
 $\rho(\bx)$ refers to the local dark matter energy density, $c=-8\pi G/m_A^2$. For ultralight dark matter, a core structure can form at the center of the Galaxy with a radius of $\sim l_c=\lambda_{dB}/(2\pi) = 1/(m_Av_{\rm vir})$.  The energy density of the core is a constant, which can be much larger than the energy density outside the core. We expect that the photon emitted by a star inside the core may have larger $\delta n_{1i}$ and $\delta n_{2i}$ since part of the worldline of the photon is inside the core. However, we will see in the following calculation that even when the star is inside the core, $\delta n_{3i}$ is still the dominant component of the total angular deflection $\delta n_i$.

\subsection{Earth outside the core, star inside the core}\label{sec:case1}
Consider the case that the star is inside the core. It is valid for $l_s<\lambda_{\rm dB}=2\pi l_c<l_e$ (with $l_e=8\,{\rm kpc}$ and $l_s$ being respectively the Earth and star's distance to the galactic center), implying $m_A\geq 10^{-23.4}\,{\rm eV}$. We suppose that at $\lambda=\lambda_b$, the photon touches the boundary of the core. Therefore we expect $\omega_0|\lambda_s-\lambda_b|\sim l_c$. First we calculate $\delta n_{1i}$. We separate $\delta n_{1i}$ into two parts
\begin{align}
    \delta n_{1i}=\frac{c}{ \lambda_s}P_{ij}\int_{\lambda_b}^{\lambda_s}n^l h_{jl}(\lambda^{\prime}) \mathrm{d}\lambda^{\prime}+\frac{c}{ \lambda_s}P_{ij}\int_{0}^{\lambda_b}n^l h_{jl}(\lambda^{\prime}) \mathrm{d}\lambda^{\prime}\; .
\end{align}
Since the $h_{ij}$ inside the core is significantly larger,  we only calculate the first integration. In addition, the integral length is approximately one coherent length, so we can use Eq. (\ref{eq:hij-express}) to express $h_{ij}$. Insert Eq. (\ref{eq:hij-express}) into $\delta n_{1i}$, we obtain
\begin{align}
    \delta n_{1i}^{\rm DP}=\frac{c}{ \lambda_s}P_{ij}n^l \mathcal{H}_{jl}^{\rm DP}\int_{\lambda_b}^{\lambda_s}\rho(\lambda^{\prime})\cos(2m_A\omega_0 \lambda^{\prime}+2\alpha) \mathrm{d}\lambda^{\prime}\sim P_{ij}n^l \mathcal{H}_{jl}^{\rm DP}\frac{c\rho_c}{ m_A \omega_0 \lambda_s}
    \label{eq:DPDMn1}\;,
\end{align}
where $\rho_c$ refers to the energy density of the soliton core. For simplicity, we use the NFW profile to estimate $\rho_c$. Suppose the energy density of DPDM near the Earth is $\rho_0$. 
The NFW profile takes the form as (\cite{mcmillan2011mass})
\begin{align}
    \rho_h=\frac{\rho_{h,0}}{x(1+x)^2}\;,
\end{align}
where $x=r/r_h$ and $r_h=17\mathrm{kpc}$. We use the energy density near $r=l_c$ to estimate the energy density of the core. Then the ratio of $\rho_c$ and $\rho_0$ is 
\begin{align}
    \frac{\rho_c}{\rho_0}\sim\frac{l_e}{l_c}\sim l_e m_A v_{\rm vir}\;.
\end{align}
Using the expression of $\rho_c$, we obtain
\begin{align}
    \delta n_{1i}^{\rm DP}\sim P_{ij}n^l \mathcal{H}_{jl}^{\rm DP}\frac{c\rho_0 v_{\rm vir}  l_e}{ x_s}\sim cP_{ij}n^l \mathcal{H}_{jl}^{\rm DP} v_{\rm vir} \rho_0\sim v_{\rm vir} \delta n_{3i}^{\rm DP}\label{eq:compare-1}\;,
\end{align}
$x_s$ is the distance from the star to the observer, $\delta n_{3i}^{\rm DP}=-\frac{1}{2}P_{ij}h_{ jk}^{\rm DP}(0)n^{k}$.  $l_e/x_s$ is an order one parameter, so we ignore it. The virial velocity is $v_{\rm vir}\sim 10^{-3}$. Therefore, $\delta n_{1i}$ is suppressed compared to $\delta n_{3i}$.

Next we consider $\delta n_{2i}$. We exchange the order of integration in $\delta n_{2i}$, which yields
\begin{align}
    \delta n_{2i}^{\rm DP}&=\frac{\omega_0}{ 2\lambda_s}P_{ij}\int_{0}^{\lambda_s}\mathrm{d}\lambda^{\prime \prime}\int_{\lambda^{\prime \prime}}^{\lambda_s}\mathrm{d}\lambda^{\prime} n^l n^k \partial_j h_{kl}^{\rm DP}(\lambda^{\prime\prime}) \nonumber \\
    &=\frac{\omega_0}{ 2\lambda_s}P_{ij}\int_{0}^{\lambda_s}\mathrm{d}\lambda^{\prime \prime}(\lambda_s-\lambda^{\prime\prime}) n^l n^k \partial_j h_{kl}^{\rm DP}(\lambda^{\prime\prime}) \nonumber \\
    &=\frac{\omega_0}{ 2\lambda_s}P_{ij}\int_{\lambda_b}^{\lambda_s}\mathrm{d}\lambda^{\prime \prime}(\lambda_s-\lambda^{\prime\prime}) n^l n^k \partial_j h_{kl}^{\rm DP}(\lambda^{\prime\prime})+\frac{\omega_0}{ 2\lambda_s}P_{ij}\int_{0}^{\lambda_b}\mathrm{d}\lambda^{\prime \prime}(\lambda_s-\lambda^{\prime\prime}) n^l n^k \partial_j h_{kl}^{\rm DP}(\lambda^{\prime\prime})\;.
\end{align}
Again we only calculate the first integral. Inside the core, we find $\omega_0|\lambda_s-\lambda^{\prime\prime}|\sim l_c=\frac{1}{m_A v_{\rm vir}}$ and $\partial_i h_{jk}^{\rm DP}\sim m_A v_{\rm vir} h_{jk}^{\rm DP}$. Then we have
\begin{align}
    \delta n_{2i}^{\rm DP}&=\frac{\omega_0}{ 2\lambda_s}P_{ij}\int_{\lambda_b}^{\lambda_s}\mathrm{d}\lambda^{\prime \prime}(\lambda_s-\lambda^{\prime\prime}) n^l n^k \partial_j h_{kl}^{\rm DP}(\lambda^{\prime\prime}) \nonumber\\
    &\sim \frac{1}{ 2\lambda_s}P_{ij}\int_{\lambda_b}^{\lambda_s}\mathrm{d}\lambda^{\prime \prime}\left(\frac{1}{m_Av_{\rm vir}}\right) n^l n^k m_Av_{\rm vir} h_{kl}^{\rm DP}(\lambda^{\prime\prime}) \nonumber \\
    &\sim  \frac{1}{ 2\lambda_s}P_{ij}\int_{\lambda_b}^{\lambda_s}\mathrm{d}\lambda^{\prime \prime} n^l n^k  h_{kl}^{\rm DP}(\lambda^{\prime\prime})\sim \delta n_{1i}^{\rm DP}\;, \label{eq:DPDMn2}
\end{align}
where we find $\omega_0(\lambda_s-\lambda^{\prime\prime})$ and the spatial derivative of $h_{ij}^{\rm DP}$ cancel with each other, making $\delta n_{1i}^{\rm DP}$ and $\delta n_{2i}^{\rm DP}$ at the same order. Then similar to $\delta n_{1i}^{\rm DP}$, $\delta n_{2i}^{\rm DP}$ is also suppressed compared to $\delta n_{3i}^{\rm DP}$. 

\subsection{Both the Earth and the star are inside the core}\label{sec:case2}
We assume that the star and the Earth are inside the core. It requires $m_A< 10^{-23.4}$ to include the Earth inside the core. We treat the dark matter density at the core as a constant $\rho(\bx) = \rho_c$. Then the first term $\delta n_{1i}$ in Eq.~(\ref{eq:three_dn}) has the approximated value
\begin{align}
    \frac{\delta n_{1i}^{\rm DP}}{\delta n_{3i}^{\rm DP}}&\leq
    \frac{l_{\rm Comp}}{D}\sin\left( \frac{D}{l_{\rm Comp}}\right)\sim
    \begin{cases}
    l_{\rm Comp}/D\,, &l_{\rm Comp}\ll D\\
          1\,,&l_{\rm Comp}\gg D
    \end{cases}\,,
\end{align}
where $\omega_0\lambda_s=D$ is the distance of the star, $1/m_A = l_{\rm Comp}$ is the Compton wavelength of the dark photon field. We find that $\delta n_{1i}$ consistently smaller than $\delta n_{3i}$ when $l_{\rm Comp}\ll D$, which holds true for $m_A \geq 10^{-24}\,{\rm eV}$ with the typical star distance of $1\,{\rm kpc}$. For $\delta n_{2i}^{\rm DP}$, we also easily find that it is at the same order of $\delta n_{2i}^{\rm DP}$ for the same reason we described in the previous sub-section. 

\subsection{Both the Earth and the star are outside the core}\label{sec:case3}
At last, we consider when the star and the Earth are outside the core. It is valid for a larger dark matter mass $m_A\gtrsim 10^{-23}\,{\rm eV}$, depending on the location of the star. Without loss of generality, we assume that the star lies on the line between the Earth and the galaxy center and the boundary of the core. Then, from previous sections, we have the dark matter density at the star $\rho_s = \rho_c \simeq \rho_0 l_c/l_e $ and 
\begin{equation}
    \frac{\delta n_{1i}^{\rm DP}}{\delta n_{3i}^{\rm DP}}\lesssim \frac{l_e}{l_c}\frac{l_{\rm Comp}}{D} \sin\left(\frac{D}{l_{\rm Comp}}\right)\sim 
    \begin{cases}
        v_{\rm vir}\l_e/D\,, &l_{\rm Comp}\ll D\\
         l_e/l_c\,\,, &l_{\rm Comp}\gg D
    \end{cases}\,.
\end{equation} 
where we used $D\gg l_{\rm Comp}$, $l_c = v_{\rm vir}l_{\rm Comp}$ and $l_e\simeq 8\,{\rm kpc}$ and $D\sim 1\,{\rm kpc}$. Similarly, $\delta n_{2i}^{\rm DP}$ is at the same order of $\delta n_{1i}^{\rm DP}$. As we estimated before, with the mass $m_A\geq 10^{-24}\,{\rm eV}$, we always have $l_{\rm Comp}\ll D$. Considering that $v_{\rm vir}\simeq 10^{-3}$, it is safe to ignore $\delta n_{1i}$ and $\delta n_{2i}$.  

\bibliography{ref}
\end{document}